\documentclass[twocolumn,showpacs,preprintnumbers,superscriptaddress,aps,prb,amsmath,amssymb,floatfix]{revtex4-1}
\usepackage{txfonts}
\usepackage{amsfonts}
\usepackage{bm}
\usepackage{graphicx}
\usepackage{epstopdf}
\usepackage{color}

\def\bk{{\bf k}}

\def\bq{{\bf q}}
\def\bQ{{\bf Q}}
\def\b0{{\bf 0}}
\def\br{{\bf r}}

\def\dag{\dagger}

\def\bra{\langle}
\def\ket{\rangle}

\def\vev#1{\langle{#1}\rangle}

\def\non{\nonumber\\}

\begin{document}

\title{Umbrella-coplanar transition in the triangular XXZ model with arbitrary spin}

\author{Giacomo Marmorini}
\email[Corresponding author: ]{giacomo@yukawa.kyoto-u.ac.jp}
\affiliation{Yukawa Institute for Theoretical Physics, Kyoto University, Kyoto 606-8502, Japan}
\affiliation{Research and Education Center for Natural Sciences, Keio University, Kanagawa 223-8521, Japan}
\author{Daisuke Yamamoto}
\affiliation{Waseda Institute for Advanced Study, Waseda University, Tokyo 169-8050, Japan}
\author{Ippei Danshita}
\affiliation{Yukawa Institute for Theoretical Physics, Kyoto University, Kyoto 606-8502, Japan}


\begin{abstract}
The quantum triangular XXZ model has recently enjoyed a wealth of new theoretical results, especially in relation to the  modeling of the Ba$_3$CoSb$_2$O$_9$ compound. In particular, it has been understood that in a longitudinal magnetic field the umbrella (cone) phase, classically stable in all the easy-plane region of the ground-state phase diagram, is considerably reduced by the effect of quantum fluctuations. We provide more quantitative information for this phenomenon at arbitrary value of the site spin $S$, by employing the dilute Bose gas expansion, valid in the high field regime; our results improve the available estimates of the $1/S$ expansion. We quantify the extent to which a higher spin suppresses the effect of quantum fluctuations.  {Besides, we show how in three-dimensional layered systems a relatively small antiferromagnetic interlayer coupling has a similar consequence of bringing back the umbrella phase in some part of the phase diagram.}
\end{abstract}


\maketitle

\section{Introduction} \label{sec:intro}

Frustrated quantum {antiferromagnets} continue to provide opportunities for the investigation of new states of matter and therefore are still nowadays the object of intense experimental and theoretical effort.\citep{starykh-15} The interplay of frustration and quantum fluctuations can determine dramatic effects on the ground state of spin systems compared to the simple classical picture. The suppression of dipolar magnetic order in favor of spin liquid states\cite{balents-10,han-12} or multipolar order such as the spin nematic state\cite{andreev-84,chandra-91,chubukov-91,shannon-06} is perhaps the most striking phenomenon; however, there are also interesting situations such that  an order of magnetic moments can be stabilized in the ground state, which does not have an analog in the ground state phase diagram of related classical systems.\cite{chubukov-91-2,chen-13} Also, it is well understood that phase transition lines in frustrated systems are more subject to large quantum correction, especially due to degeneracy or quasi-degeneracy in the energy landscape.

The triangular Heisenberg antiferromagnet is perhaps the most studied theoretical model in the field. A lot of its features have been understood over the years, including the removal of classical degeneracies by quantum or thermal fluctuations and its signature one-third magnetization plateau.\cite{chubukov-91-2,sakai-11,seabra-11}  One of the most natural extensions of this model consists in the introduction of anisotropy in spin space, in the form of XXZ anisotropy,\cite{nishimori-86,yoshikawa-04} namely a difference between the $XY$ coupling $J$ and the Ising coupling $J_z$ [see Eq.~(\ref{hamiltonian})], which is in general induced in real materials by a combination of crystal field and spin-orbit coupling. Recently there has been a remarkable revival of the triangular XXZ model, especially due to very interesting experimental works on Ba$_3$CoSb$_2$O$_9$.\cite{shirata-12,zhou-12,susuki-13,koutroulakis-13} This compound is particularly attracting because it features a stack of geometrically almost perfect triangular lattices made of Co$^{2+}$ ions, which in the presence of trigonal crystal field and spin-orbit coupling  carry an effective spin $S=1/2$ at temperatures well below the scale set by the spin-orbit coupling,\cite{lines-63} in this case about 250 K. Besides, the highly symmetric crystal structure forbids Dzyaloshinsky-Moriya interaction up to the third-nearest neighbor in the plane and between any pair of spins along the $c$ axis. {Magnetic ordering is observed below 3.8 K}. Electron spin resonance\cite{susuki-13} and, very recently, inelastic neutron scattering\cite{ma-15} experiments also provided evidence for XXZ anisotropy of the easy-plane type, as well as a small nearest-neighbor interlayer exchange coupling (a few percent of the intralayer one). Currently Ba$_3$CoSb$_2$O$_9$ is believed to be a very good physical realization of the XXZ model on a vertically stacked triangular lattice (the importance of the interlayer coupling, however small, has been highlighted in {Refs.~\onlinecite{yamamoto-15,yamamoto-15-2}}); this is strongly supported by the agreement between the magnetization process in longitudinal and transverse magnetic field and the model calculation.\cite{yamamoto-14,yamamoto-15} However, we can not exclude that a refinement of the theoretical model should be pursued to explain further experimental results.\cite{koutroulakis-13,ma-15}

{This line of research is being expanded to other compounds with identical or very similar structure, such as  Ba$_3$NiNb$_2$O$_9$ ($S=1$)\cite{hwang-12,sun-15}, Ba$_3$CoNb$_2$O$_9$ ($S=1/2$)\cite{lee-14,sun-15}, Ba$_3$MnNb$_2$O$_9$ ($S=5/2$)\cite{tian-14,sun-15} and Ba$_3$NiSb$_2$O$_9$ ($S=1$).\cite{shirata-11} While in general all these materials feature magnetic ions occupying the sites of triangular lattice layers which are arranged in a simple AA type of stacking, it becomes interesting to understand how various details of the single material, such as the value of $S$, the dimensionality (strength of the interlayer coupling), certain spin anisotropies, etc., influence the actual observations.}

\begin{figure}[tb]
\includegraphics[width=\columnwidth]{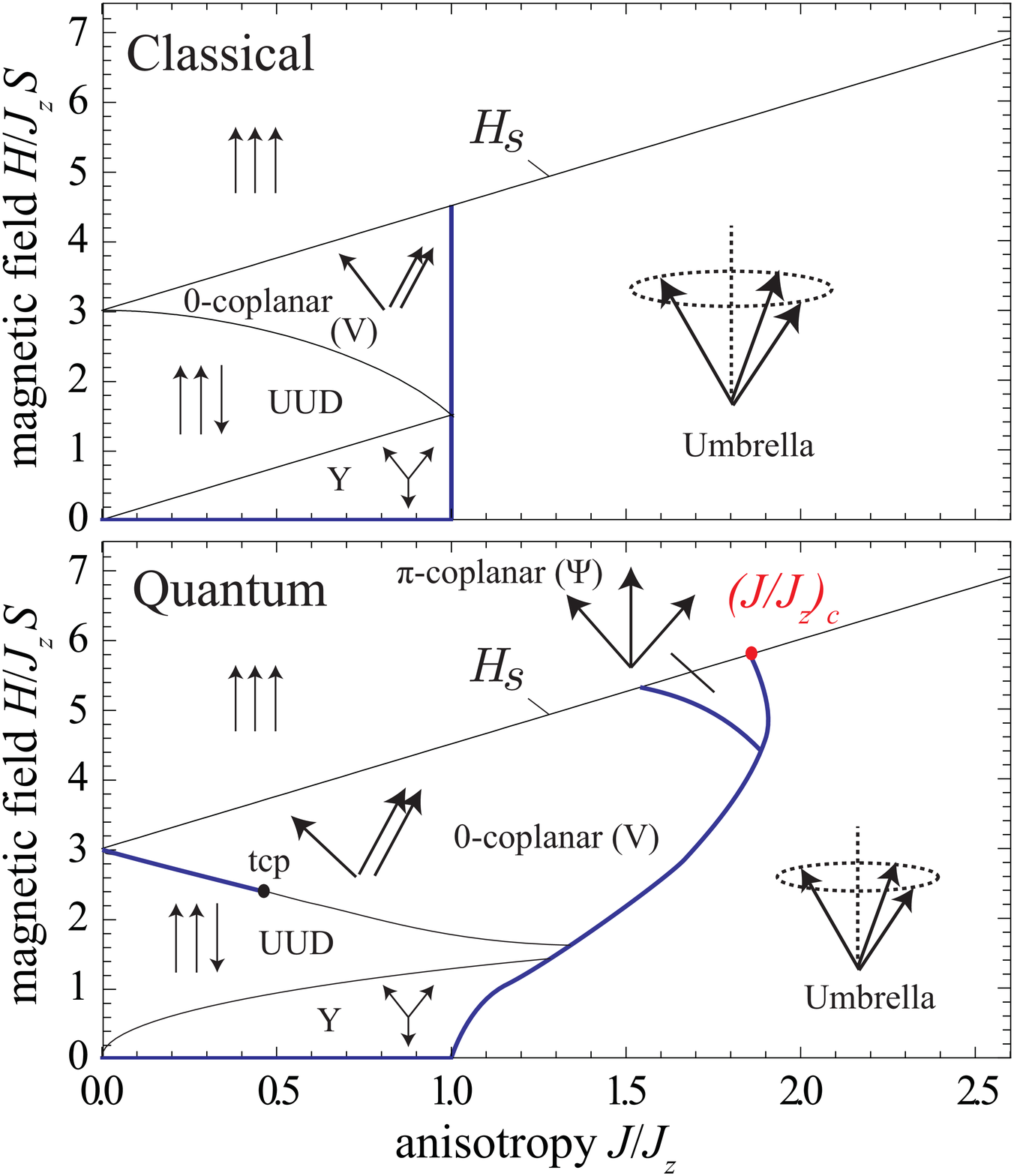}
\par \bigskip
{
\begin{tabular}{c | c c c c c}
$S$ & 1/2  & 1 & 3/2 & 2 & 5/2 \\
\hline
$(J/J_z)_c$  & 2.218 & 1.554 &  1.361 & 1.269 & 1.214
\end{tabular}
}
\caption{(Top) Classical phase diagram of the {two-dimensional} triangular XXZ model in a longitudinal magnetic field. {``UUD'' stands for up-up-down (or 1/3-magnetization plateau) state. Black and blue lines denote continuous and discontinuous phase transitions respectively. (Middle) Sketch of the same phase diagram as expected after the inclusion of quantum fluctuations for some finite value of the spin $S$; in particular, the umbrella-coplanar  transition point at saturation, $(J/J_z)_c$, is shifted to the easy-plane region by quantum fluctuations. ``tcp'' stands for tricritical point. (Bottom) Quantitative determination of $(J/J_z)_c$ for the lowest values of $S$ as obtained in Sec.~\ref{sec:results}.}}
\label{classquantpd}
\end{figure}

The latest theoretical works\cite{yamamoto-14,sellmann-14,yamamoto-15} thoroughly mapped out the magnetic phase diagram of the spin 1/2 triangular XXZ model, pointing out numerous interesting features (see Fig.~\ref{classquantpd}). Among these, in the case of {the purely two-dimensional model in} longitudinal applied field, the extension of the one-third magnetization plateau phase to the easy-plane region up to $J/J_z\simeq 1.4$ (or more according to Ref.~\onlinecite{sellmann-14}); the possibility of a new coplanar phase (named $\pi$-coplanar or $\Psi$) emerging from a peculiar lifting of the classical ground-state degeneracy; the presence of a tricritical point along the transition line between the plateau and the high field $0$-coplanar (or V) state, which changes from second to first order going towards the Ising limit at about  $J/J_z\simeq 0.44$. {In the case of layered systems in a transverse applied field,} the appearance of a new first-order phase transition in the easy-plane region at about 70\% of the saturation magnetization, which is induced by any small but non-vanishing antiferromagnetic interlayer exchange coupling.

In case of larger but finite spin, $S>1/2$, it is naturally expected that the effect of quantum fluctuation will be milder, although it is in general difficult to quantify this expectation and  not many studies can be found in literature; besides, a considerable part of them concentrates on the isotropic (Heisenberg) model. In fact, for this model there is quite a clear understanding of how the one-third magnetization plateau shrinks as a function of $S$. \cite{shirata-11,gotze-16,takano-11,zhitomirsky-15} Other studies based on the $1/S$ expansion\cite{starykh-14, nikuni-95} give us information about the generically anisotropic model in a longitudinal high magnetic field and find that the phase diagram in that region is qualitatively the same as the one found in Ref.~\onlinecite{yamamoto-14} for $S=1/2$. On these grounds and assuming some more analogy with the $S=1/2$ case, in Fig~\ref{classquantpd} (middle panel)  we draw a sketch of how the phase diagram of the purely two-dimesional model in longitudinal field should look like for a fixed finite spin $S$. Such a sketch is meant to illustrate the picture that is currently available, but clearly it should be taken with a certain amount of caution. In particular we remind that the tricritical point in the easy-axis region has been confirmed so far only for $S=1/2$, so that it can either disappear continuously for $S\to \infty$ or at some critical value of $S$. Also, let us mention that Ref.~\onlinecite{sellmann-14}, based on  exact diagonalization of the $S=1/2$ model, claims that the $\pi$-coplanar state corresponds to a spurious phase which is present in finite-size calculations and disappears in the thermodynamic limit; however, a more thorough analysis of the exact diagonalization data in the three-magnon sector shows that the spurious phase is actually a chirally symmetric combination of finite-size umbrella states, while the lowest eigenstate in the coplanar region is always doubly degenerate.\cite{yamamoto-16}

{Here we want to focus on a specific feature, namely the shift of the umbrella-coplanar transition line in a longitudinal magnetic field $H$, and treat it quantitatively.} The transition in question classically sits at the isotropic line $J=J_z$ (see Fig.~\ref{classquantpd}) and is discontinuous but with no hysteresis, due to the classical continuous ground-state degeneracy of the Heisenberg model. As explicitly seen in the $S=1/2$ case,\cite{yamamoto-14} quantum fluctuations make it {hysteretic} and shift it towards the easy-plane side [except for the SU(2)-symmetric point ($J=J_z, H=0$)]. To calculate this shift for any strength of the applied field is a difficult task, and to our knowledge this has been achieved only for $S=1/2$. The problem for arbitrary spin $S$ has been approached with a Holstein-Primakoff mapping in the high-field regime, namely near saturation; in this case the reference state is the fully polarized state (vacuum of the Holstein-Primakoff bosons), which is an exact eigenstate of the Hamiltonian and is thus protected from quantum fluctuations. The transition point in the vicinity of the saturation field $H_S$ is determined at leading order in $1/S$ as\cite{starykh-14}
\begin{eqnarray}
 \left(\frac{J}{J_z}\right)_{c,1/S}= \left( 1-\frac{0.53}{S} \right)^{-1}.
 \label{oneovers}
\end{eqnarray}
In this paper, we will employ a different strategy, based on the fact that near saturation the Holstein-Primakoff bosons are dilute. This implies that it is appropriate to consider an expansion in the boson number (per site) while treating the $1/S$ series exactly, by means of a {normal ordering}. This idea is already present in Ref.~\onlinecite{batyev-84} and leads to a dilute Bose gas treatment \`a la Beliaev,\cite{beliaev1958application,fetter-71,abrikosov-75} which is exact in the limit $H \to H_S$. We thus obtain a quantitatively precise result for the transition point $(J/J_z)_c$ just below the saturation field for any value of the spin $S$. 
{Furthermore, having in mind the application to real materials, after the purely two-dimensional model, we analyze how three-dimensionality, in the form of triangular lattice layers interacting via an  interlayer exchange coupling, can in turn influence the location of the transition point. We note that in our framework it is technically difficult, even if it is possible in principle, to address the distinction between $0$-coplanar and $\pi$-coplanar states (see Sec.~\ref{sec:setup} for a brief explanation of this point); thus we do not attempt to do that here, where we refer to the coplanar phase in general. The umbrella-coplanar transition problem is indeed well-defined independently of such a distinction.}

The paper is organized as follows. In  Sec.~\ref{sec:setup}, we set up the problem and introduce the theory of dilute interacting bosons that we will use to address it. In Sec.~\ref{sec:bseq}, some details about the calculation of quantum effects on the ground state of the model are given, with particular attention to the method that we use to deal with infrared divergences that show up in two dimensions. Sections~\ref{sec:results} and \ref{sec:concl} are devoted to the presentation of the results and a the conclusions, respectively.

\section{The triangular XXZ model in a magnetic field and its mapping to a dilute Bose gas} \label{sec:setup}

\begin{figure}[tb]
\includegraphics[width=0.9\columnwidth]{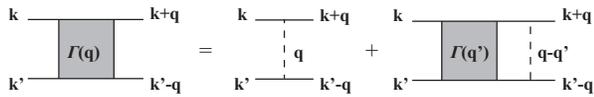}
\caption{Diagrammatic representation of the
Bethe-Salpeter equation (\ref{bseq}).
The filled squares and the dashed lines represent the renormalized and the bare interaction respectively.}
\label{ladder}
\end{figure}

The Hamiltonian of the triangular XXZ model in a longitudinal magnetic field is given by
{
\begin{eqnarray}
\hat{\mathcal{H}}&=&
J\!\sum_{\langle i,j\rangle}\!\Big(\hat{S}_i^x\hat{S}_j^x+\hat{S}_i^y\hat{S}_j^y\Big)\!+\!J_z\!\sum_{\langle i,j\rangle}\hat{S}_i^z\hat{S}_j^z\!-\!H\!\sum_{i}\hat{S}^z_{i} \non
&& + J^\perp \!\sum_{\langle i,j\rangle_\perp} \!\left(\hat{S}_i^x\hat{S}_j^x+\hat{S}_i^y\hat{S}_j^y\right)\!+\!J^\perp_z\!\sum_{\langle i,j\rangle_\perp} \hat{S}_i^z\hat{S}_j^z.
\label{hamiltonian}
\end{eqnarray}
The second line represents a nearest-neighbor interlayer coupling, since we intend to study both the purely two-dimensional system ($J^\perp,J^\perp_z=0$) and the three-dimensional layered system ($J^\perp,J^\perp_z>0$).}
We consider this model for arbitrary value of the site spin $S$. It is convenient to introduce the Holstein-Primakoff transformation
\begin{eqnarray}
& \hat{S}^+_i= \sqrt{2S}\sqrt{1-\frac{\hat{a}_i^\dag \hat{a}_i}{2S}} \,\hat{a}_i, \non
& \hat{S}^-_i =\sqrt{2S} \hat{a}_i^\dag \sqrt{1-\frac{\hat{a}_i^\dag \hat{a}_i}{2S}}, \\
& \hat{S}^z_i=S - \hat{a}_i^\dag \hat{a}_i,  \nonumber
\end{eqnarray}
to express the spin operators in terms of boson creation and annihilation operators. We want to focus on the region of high applied magnetic field, just below the saturation field, where the boson gas is dilute. In order to do so we will expand the Hamiltonian (\ref{hamiltonian}) up to four boson operators, by carefully taking into account normal ordering. Let us note that
\begin{eqnarray}
 \sqrt{1-\frac{\hat{a}_i^\dag \hat{a}_i}{2S}} = 1+\sum_{n=1}^{\infty} A_n \left( - \frac{\hat{a}_i^\dagger \hat{a}_i}{2S}
 \right)^n, \\
 A_n = \frac{1}{n!}  \prod_{j=1}^n  \left( \frac{3}{2} -j \right), \quad n\ge 1,
\end{eqnarray}
and, by virtue of $[\hat{a},.\hat{a}^\dagger]=1$,
\begin{eqnarray}
 (\hat{a}^\dagger \hat{a} )^n=\hat{a}^\dagger \hat{a} + (2^{n-1}-1)\hat{a}^\dagger \hat{a}^\dagger \hat{a}\hat{a} + \hbox{higher order in $\hat{a},\hat{a}^\dagger$}
\end{eqnarray}
for $n\ge 1$. Finally, we have
\begin{eqnarray}
  \sqrt{1-\frac{\hat{a}_i^\dag \hat{a}_i}{2S}} &= & 1+ \sum_{n=1}^{\infty} A_n \left( - \frac{1}{2S} \right)^n \hat{a}_i^\dagger \hat{a}_i+ \ldots \non 
  & \simeq & 1 + \left( \sqrt{1-\frac{1}{2S}} -1 \right) \hat{a}_i^\dagger \hat{a}_i,
\end{eqnarray} 
where in the last line we neglected higher orders in the $a,a^\dagger$ operators.\cite{batyev-84} Note, however, that we have not taken any truncation in $1/S$.
After substituting and going to momentum space the Hamiltonian (\ref{hamiltonian}) becomes, up to a constant,
\begin{eqnarray}
\hat{\mathcal{H}} &= &\sum_{\bk} [2S \epsilon(\bk) -\mu]\, \hat{a}^\dag_{\bk }\hat{a}_{\bk}  \non 
&& + \frac{1}{2M} \sum_{\bk,\bk',\bq} { V(\bq;\bk,\bk') } \, \hat{a}^{\dag}_{\bk+\bq} \hat{a}^{\dag}_{\bk'-\bq} \hat{a}_{\bk' } \hat{a}_{\bk},
\label{boseham}
\end{eqnarray}
where $M$ is the number of lattice sites and
\begin{eqnarray}
\epsilon (\bk) & =& J  \left( \frac{3}{2}+ \nu(\bk) \right) + J^\perp \cos k_z+|J^\perp|, \\
\nu(\bk) &=& \cos ({k_x}) + \cos \left(\frac{k_x}{2}+\frac{\sqrt{3}
   k_y}{2}\right) \non 
   && +\cos
   \left(\frac{{k_x}}{2}-\frac{\sqrt{3}
   {k_y}}{2}\right), \\
V(\bq;\bk,\bk')&=& V^\parallel(\bq) + \frac{V^\perp(\bk+\bq) + V^\perp(\bk'-\bq)}{2} \non 
   && + \frac{V^\perp(\bk) + V^\perp(\bk')}{2}, \\
V^\parallel(\bq) &=&  2 J_z \, {\nu(\bq)} +2J^\perp_z  \cos q_z,\\
V^\perp (\bk) &=&  4S \left( \sqrt{1-\frac{1}{2S}} -1 \right) ( J \nu(\bk) +J^\perp  \cos q_z)\non &\equiv & 4S {(K-1)} J \nu(\bk) +4S{(K-1)}  J^\perp  \cos q_z,\\
\mu &=& {6} \left( J_z+\frac{J}{2} \right)S+ 2S (J^\perp_z +|J^\perp|)- H \non &\equiv & H_S-H. \label{satf} 
\end{eqnarray}
Equation~(\ref{satf}) contains the definition of the saturation field $H_S$ {as a function of the various couplings and the spin $S$.} It represents the value of the applied field at which all the spins are polarized, or, in other words, the vacuum of bosons; this is clearly an exact eigenstate of the Hamiltonian. The inequivalent minima of the single-particle energy band are given by ${\bf k}=\pm {\bf Q}\equiv\pm(4\pi/3,0,0)$ [respectively $\pm(4\pi/3,0,\pi)$] for $J^\perp <0$ (respectively $J^\perp>0$) and we expect that at zero temperature most bosons occupy these states, in such a way that the corresponding operators acquire a non-zero expectation value, namely  $\langle \hat{a}_{\pm{\bf Q}} \rangle \equiv \psi_{\pm {\bf Q}} \neq 0$. To determine these expectation values is a difficult task due to the presence of interactions that are not weak in general. However, in the dilute limit $\mu \to 0$ ($H\to H_S$), we can employ the dilute Bose gas expansion\cite{beliaev1958application} to approach the problem. {First we operate the substitution $\hat{a}_{\pm{\bf Q}}  \to \psi_{\pm {\bf Q}} +\hat{a}_{\pm{\bf Q}}$ in Eq.~(\ref{boseham}).\cite{nikuni-95} After this, we still consider the quadratic term as in the first line of Eq.~(\ref{boseham}) as the free part of the Hamiltonian, and all the other terms as interactions. We look for the effective ground state energy; technically this is given by the generating functional of the one-particle-irreducible (1PI) correlation functions\cite{ZinnJustin:2002ru} at zero frequency. In the cumulant expansion  it is sufficient to retain  up to fourth order terms in $\psi_{\pm {\bf Q}}$, so that we can formally write}
\begin{eqnarray}
\frac{E_0}{M}&=&-\mu \left(|\psi_{{\bf Q}}|^2+|\psi_{-{\bf Q}}|^2\right) \non 
&& +\frac{{\mathit{\Gamma}}_1}{2}\left(|\psi_{{\bf Q}}|^4+|\psi_{-{\bf Q}}|^4\right)\!+\!{\mathit{\Gamma}}_2|\psi_{{\bf Q}}|^2|\psi_{-{\bf Q}}|^2.
\label{eff}
\end{eqnarray}
The parameters $\mathit{\Gamma}_1$ and $\mathit{\Gamma}_2$ are given by appropriate combinations of four-point correlation functions, namely
$\mathit{\Gamma}_1= \mathit{\Gamma}({\bf 0};\bQ,\bQ)$ and $\mathit{\Gamma}_2= \mathit{\Gamma}({\bf 0};\bQ,-\bQ)+ \mathit{\Gamma}(-2\bQ;\bQ,-\bQ)$. Here, $\mathit{\Gamma}(\bq;\bk,\bk') $ denotes a generic zero-frequency renormalized vertex (see Fig.~\ref{ladder}) as a function of two incoming and one exchanged momenta, respectively $\bk,\bk'$ and $\bq$. The calculation of this quantity will be the subject of Section~\ref{sec:bseq}. {Equation}~\eqref{eff} is minimized by a single-mode Bose-Einstein condensate (BEC) ($|\psi_{{\bf Q}}| =\mu/\mathit{\Gamma}_1$ and $|\psi_{-{\bf Q}}|= 0$ or vice versa)  for $\mathit{\Gamma}_1<\mathit{\Gamma}_2$ and by a two-mode BEC [$|\psi_{{\bf Q}}|=|\psi_{-{\bf Q}}| =\mu /(\mathit{\Gamma}_1+\mathit{\Gamma}_2)$] for $\mathit{\Gamma}_1>\mathit{\Gamma}_2$. In the spin language the single-mode BEC reads
\begin{align}
&\vev{\hat{S}^x_j}=  \sqrt{2S\frac{H_S-H}{\mathit{\Gamma}_1}}\cos\left(\bQ\cdot \br_j  \right),   \non
&\vev{\hat{S}^y_j}= \sqrt{2S\frac{H_S-H}{\mathit{\Gamma}_1}}\sin\left(\bQ\cdot \br_j\right),  \non
&\vev{\hat{S}^z_j}= S- \frac{H_S-H}{\mathit{\Gamma}_1},
\label{spiral}
\end{align}
upon a global spin rotation about the $z$ axis. This state is called umbrella or cone state in the literature. The density of bosons is related to the applied magnetic field via $\rho= \frac{H_S-H}{\mathit{\Gamma}_1}$. The two-mode BEC state corresponds to a one-parameter continuous family of coplanar states, depending on the relative phase $\phi=\arg (\psi_{{\bf Q}}/\psi_{-{\bf Q}})$. However, as explained in Ref.~\onlinecite{nikuni-95}, $\phi$ can be restricted to a finite set of values by considering the structure of the sixth-order corrections to Eq.~(\ref{eff}); moreover, by exploiting the sublattice exchange symmetry, it can be further constrained to be either $0$ or $\pi$,\footnote{More precisely the relevant correction to the ground state energy is $\delta E_0/M=2\mathit{\Gamma}_3 \psi^6 \cos 3\phi$, which is minimized by $\phi=0, 2\pi/3$, or $4\pi/3$ for $\mathit{\Gamma}_3<0$ and  $\phi=\pi/3,\pi$, or $5\pi/3$ for $\mathit{\Gamma}_3>0$. The three-fold degeneracy for fixed $\mathit{\Gamma}_3$ correspond to exchanging the three sublattices of the triangular lattice; up to such an exchange one can take $\phi$ to be either $0$ or $\pi$. It is the calculation of $\mathit{\Gamma_3}$  that requires the infinite summation of diagrams referred to at the end of Sec.~\ref{sec:setup} in the text.} from which the so-called $0$-coplanar and $\pi$-coplanar states (otherwise known as $V$ and $\Psi$ states) given by
\begin{align}
&\vev{\hat{S}^x_j}= 2 \sqrt{2S\frac{H_S-H}{\mathit{\Gamma}_1+\mathit{\Gamma}_2}}\cos\left(\bQ\cdot \br_j-\frac{\phi}{2} \right)  ,  \non
&\vev{\hat{S}^y_j}= 2 \sqrt{2S\frac{H_S-H}{\mathit{\Gamma}_1+\mathit{\Gamma}_2}}\cos\left(\bQ\cdot \br_j-\frac{\phi}{2} \right)  ,  \non
&\vev{\hat{S}^z_j}= {S}-4 \frac{H_S-H}{\mathit{\Gamma}_1+\mathit{\Gamma}_2} \cos^2\left(\bQ\cdot \br_j-\frac{\phi}{2} \right),
\label{coplanar}
\end{align}
with $\phi=0$ and $\phi=\pi$ respectively, upon a global spin rotation about the $z$ axis. Here $\rho=\frac{H_S-H}{\mathit{\Gamma}_1+\mathit{\Gamma}_2}$ represents the density of bosons at both $\bQ$ and $-\bQ$. {Within the present theory, the calculation of the sixth-order corrections that would generate an energy split between the two coplanar states requires the summation of an infinite class of Feynman diagrams\cite{nikuni-95,yamamoto-14} contributing to a three particle scattering process, which is still an open problem. However, this does not affect the treatment of the umbrella-coplanar transition, which can be addressed independently.}

\section{Bethe-Salpeter equation} \label{sec:bseq}
The calculation of the renormalized vertex $\mathit{\Gamma}(\bq;\bk,\bk') $ in the dilute limit formally amounts to solving the Bethe-Salpeter equation
\begin{eqnarray}
 &&\mathit{\Gamma}(\bq;\bk,\bk') =  V(\bq;\bk,\bk')  \non 
 & &- \frac{1}{2SM} \sum_{\bq' \in {\rm BZ} } 
 \frac{V(\bq-\bq';\bk+\bq',\bk'-\bq')}{\epsilon(\bk+\bq')+\epsilon(\bk'-\bq')} \mathit{\Gamma}(\bq';\bk,\bk'),
\label{bseq}
\end{eqnarray}
where the total energy and the chemical potential have been set to zero. {In the spin language, zero chemical potential means that for a gien set of $J,J_z,J^\perp,J^\perp_z$ and $S$ the magnetic dield $H$ is tuned to the value $H_S$ defined in Eq.~(\ref{satf}).}  The solution represents the summation of the infinite series of ladder diagrams, which are the dominant ones in the dilute limit.\cite{fetter-71,abrikosov-75,schick1971two} However, it is well-known that in two dimensions Eq.~(\eqref{bseq}) gives a logarithmically vanishing result,\cite{schick1971two,fisher1988dilute} due to the infrared singularities in the kernel. On the other hand, it is too naive to just take a zero chemical potential, even though this does not give any problem in the corresponding three-dimensional calculation. We are actually interested in the situation of a very small but non-zero chemical potential $\mu$, which in two dimensions provides a natural momentum cutoff of order $\sqrt{\mu}$. Below that scale, in fact, the bare single-particle dispersion gets dressed by many-body effects and the integration in Eq.~(\ref{bseq}) must be cut off (or, in the renormalization group language, the flow must be stopped).\cite{fisher1988dilute}
One more complication lies in the fact that $\mathit{\Gamma}_1=\mathit{\Gamma}_2$ at leading order in $|\ln \mu|^{-1}$ in the limit of very small $\mu$. There are several ways to deal with the above issues without giving up the relative simplicity of the theory;\cite{marmorini2013,lee-02} we choose to take full advantage of the Hamiltonian Eq.\eqref{hamiltonian} by initially taking a small interlayer coupling and 
eventually send it to zero,\cite{marmorini2013,yamamoto-14} looking at the evolution of the transition point of interest in the process. The reliability of the extrapolation procedure is discussed extensively in Ref.~\onlinecite{marmorini2013}. 

{As mentioned above, in the three-dimensional case (finite interlayer coupling), Eq.~(\ref{bseq}) gives a finite value of $\mathit{\Gamma}_1,\mathit{\Gamma}_2$, which can be then directly compared.}
Further technical details on the solution of the Bethe-Salpeter equation and the determination of the transition points are given in Appendix~\ref{sec:details}.

Let us conclude this section by making a remark on the dilute limit. As observed, for example, in Ref.~\onlinecite{ueda-09} in the context of spin systems (but well-known in the cold atom literature) the dilute limit is appropriately defined in terms of the scattering length in three dimensions. In our problem we have two scattering lengths, which are proportional to $\mathit{\Gamma}_1,\mathit{\Gamma}_2$, so that the dilute condition can be written as $\mathit{\Gamma}_i(\rho /\det h )^{1/3}\ll 1$, where $\det h$ is the determinant of the Hessian matrix of the energy dispersion at $\bQ$ (essentially the inverse mass matrix of the low-energy modes). Thus, as long as the $\mathit{\Gamma}_i$'s are finite (which is the case for the range of parameters that we choose in our calculation), one can always choose a magnetic field $H$ sufficiently close to $H_s$ in order for the above condition to be satisfied.

\section{Results} \label{sec:results}

\subsection{Two-dimensional model}

\begin{figure}[tb]
\includegraphics[width=\columnwidth]{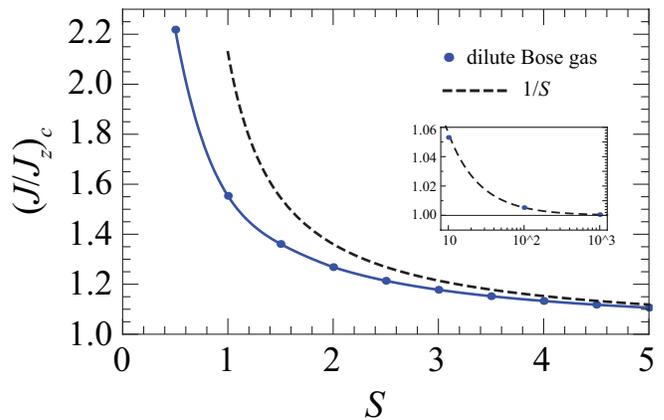}
\caption{Umbrella-coplanar transition point near saturation (blue dots) for different values of the site spin $S$ (the blue line is just a guide for the eye). {The corresponding values of the saturation field can be obtained from Eq.~(\ref{satf}) by setting $J^\perp,J^\perp_z=0$ and are listed in Table~\ref{maintab} for the first values of $S$.} The black dashed line represents the $1/S$ result,\cite{starykh-14} see Eq.~(\ref{oneovers}). The inset intends to show that the two calculations approach each other and the classical limit asymptotically for large $S$.}
\label{mainfig}
\end{figure}

The main results of our analysis are included in Fig.~\ref{mainfig}. where $(J/J_z)_c$ is plotted as a function of $S$. {The numerical results for the first spin values are explicitly displayed in Table~\ref{maintab}, together with the results of the $1/S$ theory for comparison.} As consistency checks, we note that (i) as $S\to \infty$ the transition point approaches the classical one, namely $(J/J_z)_c=1$ (see inset) and (ii) for $S=1/2$ the previous result obtained via the hard-core boson map (and consistent with the cluster mean field theory+scaling method),\cite{yamamoto-14} namely $(J/J_z)_c=2.218$. is recovered. While the former feature may appear trivial since the second term in the right-hand side (r.h.s.) of Eq.~(\ref{bseq}) is suppressed for very large spin (which, in physical terms, means that there is no renormalization since the loop diagrams vanish), the second feature is less immediate and indicates that the hard-core boson and the Holstein-Primakoff methods agree for spin 1/2,  as they should. 

The value of the transition  point decreases monotonically as $S$ increases, which is quite natural since the quantum fluctuations are expected to play a progressively weaker role. The decrease is considerably more rapid at the beginning, nevertheless the shift from the classical value is still bigger than 20\% for $S=5/2$, which is  often regarded as almost classical. 

By direct comparison, one can see that the leading order in the $1/S$ expansion in general overestimates the shift of the transition point and in particular it {is} not quantitatively accurate for $S\lesssim 3$; also, the extrapolation to $S=1/2$ is clearly not viable due to the unphysical singularity in Eq.~(\ref{oneovers}).

\begin{table}[tb]
{
\begin{tabular}{c | c c || c}
$S$ & $(J/J_z)_c$ & $(J/J_z)_{c,1/S}$ & $(H/J_z )_c/S$\\
\hline
1/2 & 2.218 & undefined & 6.327\\
1 & 1.554 & 2.128  & 5.330 \\
3/2 & 1.361 & 1.546 & 5.042 \\
2 & 1.269 & 1.361 & 4.903 \\
5/2 & 1.214 & 1.269 & 4.821 \\
3 & 1.178 & 1.215 & 4.767 \\
7/2 & 1.152 & 1.178 & 4.728 \\
\end{tabular}
\caption{Numerical result for the umbrella-coplanar transition point near saturation for the several values of the site spin $S$; in the third column we also report the result of the $1/S$ theory as calculated from Eq.~(\ref{oneovers}). The rightmost column contains the  values of the saturation magnetic field corresponding the transition points.}
\label{maintab}
}
\end{table}

\subsection{Effect of finite interlayer coupling}

\begin{figure}[tb]
\includegraphics[width=\columnwidth]{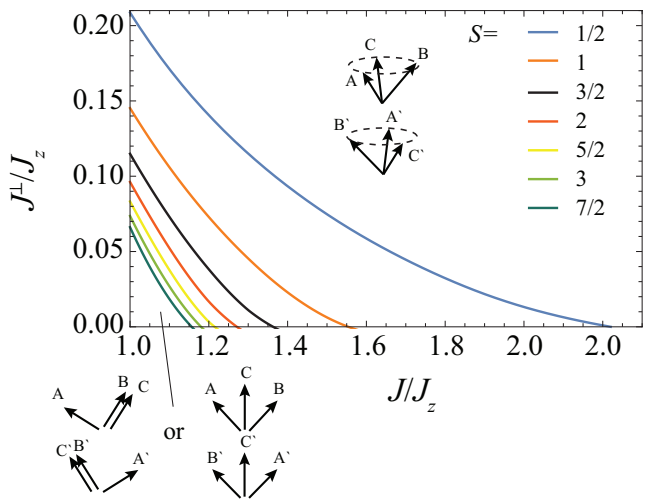}
\caption{(Top) Phase diagram in the $[J/J_z,J^\perp/J_z]$ plane near saturation for several values of the site spin $S$ {($J^\perp/J^\perp_z=1$ fixed)}.The corresponding values of the saturation field can be obtained from Eq.~(\ref{satf}). (Bottom) Zoom of the easy-plane side of the same phase diagram (corresponding to the dashed box in the upper panel). The coplanar region becomes smaller for higher $S$ and eventually disappear in the $S\to \infty$ (classical) limit.}
\label{pdanisinter}
\end{figure}

In general the interlayer exchange coupling, however small, is important to determine the ground state of layered frustrated magnets, as emphasized in Ref.~\onlinecite{yamamoto-15} for the particular case of the triangular XXZ model in a transverse field. In this section, we aim to analyze quantitatively the combined effect of antiferromagnetic interlayer coupling and {intralayer spin} anisotropy on the umbrella-coplanar transition of the triangular XXZ model in longitudinal field just below saturation. Here the umbrella and coplanar states have to be intended as associated with a well-defined stacking pattern dictated by the antiferromagnetic coupling, as depicted in Fig.~\ref{pdanisinter}, where the primed (non-primed) indices refer to the even (odd) layers. 

Since the system is three-dimensional, there are no issues related to the  infrared divergences discussed in Sec.~\ref{sec:bseq} and the calculation is essentially the same as in Ref.~\onlinecite{nikuni-95}. Here we extend it systematically to a wide range of the parameters and provide some theoretical considerations.

We have learned that quantum fluctuations in two dimensions not only select the $0$-coplanar phase in the isotropic (Heisenberg) model, but also stabilize a coplanar phase up to quite large values of easy-plane anisotropy. This effect naturally weakens for higher spin, but it is notable for all physically interesting values of $S$. One is then led to ask what is the effect of introducing three-dimensionality and an antiferromagnetic coupling in the direction perpendicular to the triangular lattice  layers.

Two facts are known about the isotropic case. On one hand in the classical limit a small interlayer coupling stabilizes the stacked umbrella state for all values of the applied field.\cite{gekht-97} On the other hand for $S=1/2$ at  $J^\perp/J\simeq 0.209$, the effect of the interlayer coupling is sufficient to contrast the quantum fluctuations and destabilize the stacked $0$-coplanar phase in favor of the stacked umbrella phase.\cite{nikuni-95} Combining these observations with the result of the previous subsection, one can argue that the phase diagram in the $[J/J_z,J^\perp/J_z]$ plane 
features a stacked coplanar phase in the lower-left corner and the stacked umbrella in the upper-right region. Furthermore the first-order phase transition line between the two progressively shifts downwards as $S$ increases, which asymptotically leads to the disappearance of the coplanar phase from the easy-plane region, consistently with the classical result. The actual calculation indeed confirms this scenario and is reported in Fig.~\ref{pdanisinter} (an isotropic interlayer coupling is used, namely $J^\perp=J^\perp_z$). The same phase diagram for the case of $S=1/2$, $J/J_z>1$,  was discussed in Ref.~\onlinecite{koutroulakis-13}. {We recall that for fixed $S$ we expect an additional first-order phase transition  between  {$0$ and  $\pi$ states} within the coplanar region,\cite{yamamoto-14,starykh-14} but its determination  goes beyond the scope of this paper (see the discussion in Sec.~\ref{sec:intro} and \ref{sec:setup}).}

\begin{figure}[tb]
\includegraphics[width=\columnwidth]{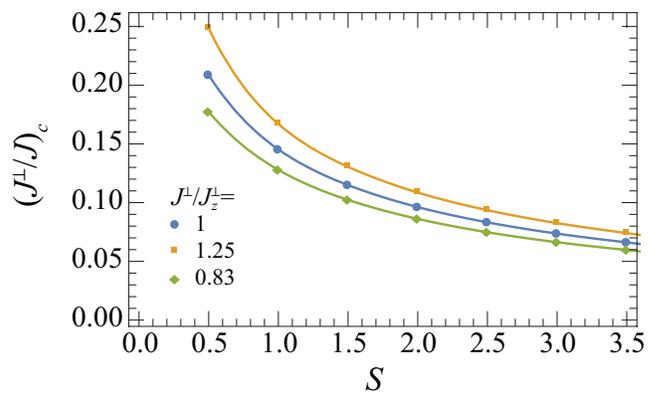}
\caption{{Model with isotropic intralayer interaction ($J=J_z$)}: value of  $J^\perp$ at which the stacked  umbrella-stacked coplanar transition takes place  as a function of the site spin $S$ (the continuous lines are just a guide for the eye) for three different value of the spin anisotropy in the direction perpendicular to the layers.}
\label{heisfig}
\end{figure}

We illustrate in a more detailed fashion the calculation for the specific case of the {model with isotropic intralayer interaction ($J=J_z$)} in Fig.~\ref{heisfig}. Additionally, we note the consequence of changing the spin anisotropy of the interlayer exchange. In general, if that anisotropy is of $XY$ type  (respectively Ising type) the coplanar region expands (respectively contracts). We have checked that this tendency occurs also when the model possesses {a (fixed) spin anisotropy} in the intralayer exchange interaction ($J\neq J_z$) .


\section{Conclusions} \label{sec:concl}

In this study, we provided an accurate determination of the umbrella-coplanar transition point of the two-dimensional triangular XXZ model near saturation for arbitrary spin $S$. This was made possible by  using the Holstein-Primakoff mapping together with {normal ordering} that allowed for the appropriate dilute boson expansion; also, the singularities of the ladder approximation in two dimensions were worked around with a sort of dimensional regularization (smooth extrapolation to zero interlayer coupling). We thus made the information that a larger $S$ weakens quantum fluctuations, and then the coplanar states, quantitative. In addition, we studied the effect of {a finite antiferromagnetic interlayer coupling (which makes the system three-dimensional),} an ingredient quite important for modeling real compounds, and found precisely how it also tends to stabilize the umbrella state. We hope that the above results can help to guide future high-field experiments on currently available and new materials based on magnetic triangular lattices.


\section*{Acknowledgments}

We thank T.~Nikuni and Y.~Kamiya for useful discussions and the anonymous referees for stimulating questions and comments. G.~M. acknowledges a research travel grant from the Mochizuki Fund of the Yukawa Memorial Foundation. This work
was partially supported by KAKENHI Grants from Japan Society for the Promotion of Science No. 25800228 (I.D.), No. 25220711 (I.D.),
and No. 26800200 (D.Y.).

\appendix
\section{Details on the Bethe-Salpeter equation} \label{sec:details}
In this section we provide some details on the solution of the Bethe-Salpeter equation. 
First  we  integrate Eq.~(\ref{bseq}) over $\bq$ and find
\begin{align}
 &\bra \mathit{\Gamma}(\bq;\bk,\bk') \ket =   \frac{V^\perp(\bk) + V^\perp(\bk')}{2} \non 
 & \quad - \frac{1}{4SM} \sum_{\bq' \in {\rm BZ}}  \frac{V^\perp(\bk+\bq') + V^\perp(\bk'-\bq') }{\epsilon(\bk+\bq')+\epsilon(\bk'-\bq')} \mathit{\Gamma}(\bq';\bk,\bk') \non 
 & =   \frac{V^\perp(\bk) + V^\perp(\bk')}{2} - (K-1) \bra \mathit{\Gamma}(\bq;\bk,\bk') \ket \non 
 & \quad +\left( \frac{3J}{2} +|J_\perp| \right) \frac{2(K-1)}{M} \sum_{\bq' \in {\rm BZ}}  \frac{ \mathit{\Gamma}(\bq';\bk,\bk')}{\epsilon(\bk+\bq')+\epsilon(\bk'-\bq')},
\label{bsequ}
\end{align}
where $\langle\ldots \rangle=(1/M)\sum_{\bq\in BZ}(\ldots)$. Restricting to the initial momenta of interest, namely $\bk,\bk'\in \{ \bQ.-\bQ\}$, the above equation reduces to 
\begin{align}
& \frac{K}{4 \kappa S(K-1)} \bra \mathit{\Gamma}(\bq;\bk,\bk') \ket \non
& + \frac{1}{2SM} \sum_{\bq' \in {\rm BZ}} \frac{\mathit{\Gamma}(\bq';\bk,\bk')}{\epsilon(\bk+\bq')+\epsilon(\bk'-\bq')} =1,
\label{bsequinf}
\end{align}
where we have defined $\kappa= -(3/2 +|J_\perp|)$. Let us introduce the two even functions $\widetilde{\mathit{\Gamma}}_1(\bq)=\mathit{\Gamma}(\bq;\bQ,\bQ)$ and $\widetilde{\mathit{\Gamma}}_2({\bq})= \mathit{\Gamma}(-\bQ+\bq;\bQ,-\bQ)+ \mathit{\Gamma}(-\bQ-\bq;\bQ,-\bQ)$. In the end $\mathit{\Gamma}_1=\widetilde{\mathit{\Gamma}}_1({\bf 0})$ and $\mathit{\Gamma}_2=\widetilde{\mathit{\Gamma}}_2(\bQ)$. By taking the ansatz
\begin{align}
\widetilde{\mathit{\Gamma}}_\alpha(\bq)=  \bra \widetilde{\mathit{\Gamma}}_\alpha \ket +  J_z A_\alpha \nu(\bq)  +J^\perp_z B_\alpha \cos {q_z}, \quad \alpha=1,2,
\label{ansatz}
\end{align}
and defining
\begin{align}
& {\bf T}(\bq) = (1,\nu(\bq), \cos q_z )^T, \\
&\tau_{ij} (\bk,\bk') = \frac{1}{2SM} \sum_{\bq'\in BZ} \frac{T_i(\bq') T_j(\bq')}
{\epsilon\left(\frac{\bk+\bk'}{2}+\bq'\right) +\epsilon\left(\frac{\bk+\bk'}{2}-\bq'\right)}, \\
&\tau^1=\tau(\bQ,\bQ), \quad \tau^2=\tau(\bQ,-\bQ),
\label{tau}
\end{align}
Eqs.~\eqref{bsequinf} and \eqref{bseq} [after the substitution Eq.~\eqref{bsequ}] can be reduced to the linear algebraic systems
\begin{widetext}
\begin{align}
& \left(
\begin{array}{ccc}
\tau^1_{11} + \frac{K}{4 \kappa S(K-1)} & J_z  \tau^1_{12} & J^\perp_z \tau^1_{13} \\
\frac{2}{3} \tau^1_{21} + \frac{KJ}{2 \kappa J_z} &  1+ \frac{2}{3} J_z  \tau^1_{22} & \frac{2}{3} J^\perp_z \tau^1_{23} \\
2\tau^1_{31} - \frac{KJ^\perp \cos (Q_z)}{ \kappa J^\perp_z}  & 2  J_z  \tau^1_{32} & 1+  2 J^\perp_z\tau^1_{33} 
\end{array}
\right)
\left(
\begin{array}{c}
\langle \widetilde{\mathit{\Gamma}}_1 \rangle \\
A_1 \\
B_1\\
\end{array}
\right)
=\left( \begin{array}{c}
   1\\
   2\\
   2
  \end{array} \right),
 \\
 & \left(
\begin{array}{ccc}
\tau^2_{11} + \frac{K}{4 \kappa S(K-1)} & J_z  \tau^2_{12} & J^\perp_z \tau^2_{13} \\
\frac{2}{3} \tau^2_{21} - \frac{KJ}{ \kappa J_z} &  1+ \frac{2}{3} J_z  \tau^2_{22} & \frac{2}{3} J^\perp_z \tau^2_{23} \\
2  \tau^2_{31} - \frac{KJ^\perp }{ \kappa J^\perp_z} & 2   J_z  \tau^2_{32} & 1+  2 J^\perp_z\tau^2_{33} 
\end{array}
\right)
\left(
\begin{array}{c}
\langle \widetilde{\mathit{\Gamma}}_2 \rangle \\
A_2 \\
B_2\\
\end{array}
\right)
=\left( \begin{array}{c}
  2\\
   -2\\
   4\cos (Q_z)\,
  \end{array} \right).
\end{align}
\end{widetext}

\begin{center}
\begin{figure}[htb]
\includegraphics[width=\columnwidth]{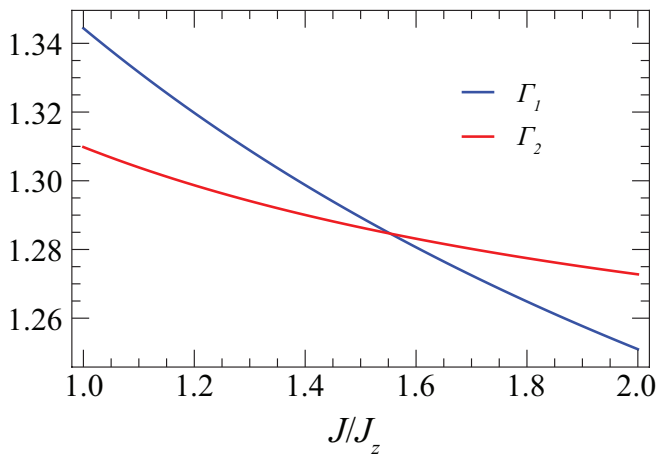}
\caption{$\mathit{\Gamma}_1,\mathit{\Gamma}_2$ for $S=1$ and $J^\perp=J^\perp_z=10^{-6}$.}
\label{cross}
\end{figure}
\end{center}

\begin{center}
\begin{figure}[htb]
\includegraphics[width=\columnwidth]{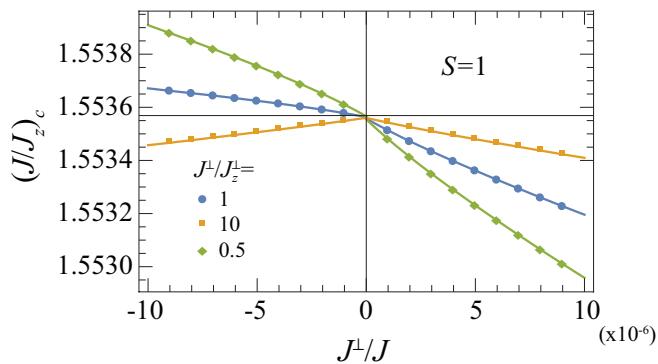}
\caption{Extrapolation of the transition point to two dimensions ($J^\perp,J^\perp_z=0$) for $S=1$ and various choices of $J^\perp/J^\perp_z$. The left (right) part represents ferromagnetic (antiferromagnetic) $J^\perp$. }
\label{extrap}
\end{figure}
\end{center}

For fixed $S,J^\perp,J^\perp_z$, it is then easy to find $\mathit{\Gamma}_1,\mathit{\Gamma}_2$ as a function of $J/J_z$. The result for one choice of the parameters is plotted in Fig.~\ref{cross}. According to the discussion in Sec.~\ref{sec:setup} the crossing point determines the location of the umbrella-coplanar transition. The next step consists in following the position of the transition point upon decreasing the interlayer coupling $J^\perp$ (down to $10^{-6}$ in our numerical calculation) and finally extrapolate it to the two-dimensional case. As shown in Fig.~\ref{extrap}, this can be done for any choice of the spin anisotropy in the interlayer direction without affecting the result, which is a further consistency check of our procedure. 

\end{document}